\documentclass[reprint,eqsecnum,prb]{revtex4-1}

\usepackage{graphicx}
\usepackage{amsmath}
\usepackage{color}
\usepackage{subfigure}

\begin{document}
%\markboth{Zhen-Gang Wang} {Area Expansion and Adsorption Isotherm in Protein Binding to Mixed Lipid Membranes}
\title{Theory of Polymer Chains in Poor Solvent: Single-Chain Structure, Solution Thermodynamics and $\Theta$ Point}
\author{Rui Wang and Zhen-Gang Wang\footnote{E-mail: zgw@caltech.edu}}

\address{Division of Chemistry and Chemical Engineering, California
Institute of Technology, Pasadena, CA 91125, USA}

\begin{abstract}
Using the language of the Flory $\chi$ parameter, we develop a
theory that unifies the treatment of the single-chain structure and
the solution thermodynamics of polymers in poor solvents. The
structure of a globule and its melting thermodynamics is examined
using the self-consistent filed theory.  Our results show that the
chain conformation involves three states prior to the
globule-to-coil transition: the \emph{fully-collapsed globule}, the
\emph{swollen globule} and the \emph{molten globule}, which are
distinguished by the core density and the interfacial thickness. By
examining the chain-length dependence of the melting of the swollen
globule, we find universal scaling behavior in the chain properties
near the $\Theta$ point. The information of density profile and free
energy of the globule is used in the dilute solution thermodynamics
to study the phase equilibrium of polymer solution. Our results show
different scaling behavior of the solubility of polymers in the
dilute solution compared to the F-H theory, both in the $\chi$
dependence and the chain-length dependence. From the perspectives of
single chain structure and solution thermodynamics, our results
verifies the consistency of the $\Theta$ point defined by different
criteria in the limit of infinite chain length: the disappearance of
the second viral coefficient, the abrupt change in chain size and
the critical point in the phase diagram of the polymer solution. Our
results show $\chi_\Theta=0.5$ (for the case of equal monomer and
solvent volume), which coincides with the value predicted from the
F-H theory.
\end{abstract}

\maketitle

\section{Introduction}
\label{sec:level1}

Flory-Huggins (F-H) theory is the cornerstone for polymer solution
thermodynamics\cite{Florybook}.  The theory assumes random mixing of
ideal chains with solvent molecules, and describes the
polymer-solvent interaction by a phenomenological parameter $\chi$.
Above the critical $\chi$ (or equivalently below the critical
temperature), the F-H theory predicts phase separation of a polymer
solution into a polymer-concentrated phase and polymer-dilute phase.
From the F-H theory, des Cloizeaux and Jannink\cite{Cloizeaux} found
that the equilibrium polymer concentration in the dilute phase well
below the critical point scales as $\ln \phi \sim -(\chi -
\frac{1}{2})^2 N$. The critical $\chi$ for polymer chains of length
$N$ is given by $\frac{1}{2}+N^{-1/2}$, and has the asymptotic value
of $\frac{1}{2}$ as $N$ approaches infinity.  It is accepted that,
in the limit of infinite chain length, the critical point for the
phase separation of the polymer solution coincides with the $\Theta$
point for the coil-to-globule transition of a single
chain\cite{deGennesbook,Freedbook,Yamakawabook,Rubinsteinbook,Fujitabook}.
Thus, $\chi_\Theta=\frac{1}{2}$ in the framework of F-H theory in
the limit of infinite chain length.

The phase diagram calculated by the F-H theory shows poor agreement
with experiments and simulation for the dilute phase and near the
critical point: the F-H theory predicts a much lower equilibrium
polymer concentration and a lower critical
$\chi$\cite{Takano,Dobashi,Madden,Yan,Panagiotopoulos}. An obvious
fact that is not captured by the F-H theory is the large localized
density fluctuation in the dilute solution. The instantaneous
picture of the solution has much higher polymer density where the
chains are located and pure solvents elsewhere, which is
significantly different from the random-mixing picture envisioned in
the F-H theory.  For polymers in poor solvents, such localized
density fluctuation takes the form of single-chain globules and
multichain clusters\cite{Grossbergbook}. By accounting for the large
localized fluctuation, we showed in previous work\cite{Wang1} that
the solubility of the polymer in the poor solvent is enhanced by
several orders of magnitude relative to the prediction of the F-H
theory -- the logarithm of the solubility scales with $N^{2/3}$
rather than $N$ as predicted from the F-H theory. The critical
$\chi$ in the phase diagram for polymers of finite chain length was
also shown to be pushed to higher values than
$\frac{1}{2}+N^{-1/2}$, consistent with computer simulation and
experiment results.

Another important fact not reflected in the F-H theory is the change
in chain conformation according to different solvent
qualities\cite{Nishio,Swislow,Wu1,Wu2,Wu3,Baysal,Hu,Binder,WQ}. With
increasing $\chi$, an isolated polymer chain will undergo the
coil-to-globule transition from the swollen coil ($R_g \sim
N^{3/5}$) in good solvent to the ideal coil ($R_g \sim N^{1/2}$) at
the $\Theta$ condition, and further to the collapsed globule ($R_g
\sim N^{1/3}$) as the solvent becomes poorer.  The coil-to-globule
transition is a unique property of macromolecules that is different
from the small molecules: even one macromolecule can form a
mesoscopic phase\cite{Williams,Grossbergbook,Lifshitz,Khokhlov}. It
is a fundamental problem relevant to many interesting phenomena,
such as protein folding\cite{Shakhnovich}, DNA
packing\cite{Gelbart}, gel network collapse\cite{Tanaka}, and
interpolymer complexation\cite{Wu4}. The coil-to-globule transition
of a flexible polymer is of second-order\cite{Lifshitz,deGennes}: in
the limit of infinite chain length, there is a well-defined
transition temperature known as the $\Theta$ point; whereas for the
chain of finite length, the transition occurs within the $\Theta$
region of the width proportional to $N^{-1/2}$.  De
Gennes\cite{deGennes} pointed out that the $\Theta$ point is a
tricritical point, so mean-field theory can be applied except for
logarithmic corrections\cite{Grassberger,Withers}. For polymer
chains with sufficient stiffness, the coil-to-globule transition is
predicted to become first-order\cite{Grosberg1,WQ}.

Given that the F-H theory provides a poor description of polymers in
poor solvent, on consideration of both the over-simplified physical
picture in the theory and the demonstrated inconsistency with
experiments and simulation, it is natural to question the validity
of the $\Theta$ value of $\chi$ predicted by the F-H theory. The
$\chi$ parameter is a general language for describing single chains,
polymer solutions, blends and block copolymers. The value of
$\chi_\Theta$ is crucially important in polymer science and
engineering, because it serves as the benchmark to categorize
different polymer-solvent pairs. However, existing theories for
coil-to-globule transition are usually couched in the language of
either the viral
coefficients\cite{Grosberg1,Grosberg2,Grosberg3,Grosberg4} or the
expansion factor\cite{deGennes,Ptitsyn,Birshtein}, which makes it
difficult to directly connect the single chain behavior to the
solution properties.  Furthermore, the second virial coefficient
used in the study of the single-chain behavior in previous work was
obtained from the F-H free energy of the homogeneous bulk phase; the
connection between this second virial coefficient and the effective
two-body interaction at the level of the single-chain Hamiltonian
has not be elucidated. In addition, the consistency of the $\Theta$
point defined by the different criteria concerning different
properties -- the second viral coefficient in dilute polymer
solution, the single-chain conformation, and the critical point in
the phase diagram of polymer solution in the limit of infinite chain
length -- has not been theoretically verified. On the other hand,
computer simulation of the $\Theta$ point is challenging due to the
limitation of chain length\cite{Binder,Bruns,Sheng} and numerical
accuracy.  On the experimental side, there are ambiguities in the
structure of single-chain globule.  For example, by measuring the
temperature dependence of the chain size for an isolated polymer in
aqueous solution, Wu and Wang\cite{Wu1,Wu2} suggested that the chain
conformation before the globule-to-coil transition involves two
globule states: a \emph{fully-collapsed globule} and a \emph{molten
globule}, where, in their words, quantitative description of the
molten globule is still a challenge. They also found it surprising
that a fully-collapsed globule still contains $66\%$ solvent inside
its volume.

The globule state of polymers has been theoretically studied
extensively by Grosberg and
coworkers\cite{Grossbergbook,Grosberg1,Grosberg2,Grosberg3,Grosberg4,Witelski}.
Using the Lifshitz theory\cite{Lifshitz}, these authors have
systematically elucidated the globule structure, globule-to-coil
transition, globule-globule interaction and the phase equilibrium in
dilute solution. The Lifshitz theory assumes ground-state dominance
and uses a virial expansion of the local interactions. For both
these two assumptions to be satisfied, the globule must on one hand
have a sharp interface and on the other hand have low monomer volume
fraction in the core.  The combination of these two assumptions
excludes a significant portion of the parameter space in the globule
state.  In addition, the globule structure in a poor solvent and the
melting thermodynamics of a globule near the $\Theta$ point are
studied by using two different theoretical formulations, the former
by using the Lifshitz theory and the later by using Flory expansion
factor.  To complete the knowledge on the globule state of polymers,
it is desirable to develop a theory that can describe the structural
change of a globule in the full parameter space of the poor solvent,
and can unify the globule structure and the globule melting in a
single theoretical framework.  In this regard, we note that Ref.
\cite{Witelski} showed that the expansion factor calculated using
the F-H free energy has a large discrepancy with that obtained using
the virial expansion. However, the density profile of the globule
and the surface energy or its connection to solubility were not
examined.

In an earlier paper, we developed a new theory in the language of
the F-H $\chi$ parameter that can unify the description of  the
single chain structure and the solution thermodynamics for polymers
in poor solvent\cite{Wang1}.  There, we focused on the equilibrium
cluster distribution, the solubility limit, and nucleation in the
supersaturated state.  In the current work, we apply the theory to
examine in detail the structure and properties of a single-chain
globule. In our theory, we employ the full self-consistent field
theory (SCFT) and treat the solution as incompressible, thus
allowing us to describe both a high-density globule and melting of
the globule as the $\Theta$ point is approached.  From the
chain-length dependence of globule melting characteristics, we
investigate the scaling behavior and determine the value of
$\chi_\Theta$ by extrapolating to the limit of infinite chain
length.  Because our theory is capable of describing the phase
equilibrium in dilute polymer solutions, we also determine
$\chi_\Theta$ from the solubility of polymers based on the intrinsic
connection between the single chain structure and solution property.

\section{Theory}

The system of polymers in poor solvent is a collection of globules
and clusters. We study this structure of large localized density
fluctuation by focusing on a subvolume of the entire solution that
contains only one globule or one multi-chain
cluster\cite{Wang1,Wangjiafang1,Wood}. The density profile and free
energy of the globule and clusters are obtained by applying
self-consistent field theory (SCFT) in the subvolume. This
information is then used in the framework of dilute solution
thermodynamics to reconstruct the solution behavior of the full
system. To distinguish between the composition within a single
globule/cluster and the overall bulk composition in the solution, we
use the notation $\rho$ and $\phi$, respectively, to denote the
volume fraction of polymer in the globule/cluster and in the
solution.

\subsection{Self-consistent Field Theory for an isolated globule/cluster}

We consider a subvolume $V$ consisting of $m$ polymer chains and $n$
solvent molecules. $m=1$ specifies the single-chain globule. We
treat the subvolume as a semi-canonical ensemble: the number of
polymers in the subvolume is fixed, whereas the solvent is connected
with a reservoir of pure solvent outside that maintains chemical
potential $\mu_s$. (Here $\mu_s$ is defined relative to that of the
pure solvent.) The polymers are assumed to be Gaussian chains of $N$
Kuhn segments (with Kuhn length $b$). For simplicity, the volume of
the solvent molecule and the chain segment are assumed to be the
same $v$. The local polymer-solvent interaction is described by the
Flory $\chi$ parameter, and the solution is treated as
incompressible.

The semi-canonical partition function can be written as:
\begin{align}
\Xi  = &\sum_{n=0}^{\infty}\frac{e^{\beta \mu_s n}}{m ! n ! v ^{Nm+n}} \prod_{j=1}^{m} \int \hat{D} \{ {\bf{R}}_j \} \prod_{k=1}^{n} \int d {\bf{r}}_k   \nonumber\\
& \delta \left(\hat{\rho}_p + \hat{\rho}_s -1 \right)  \exp \left[
-\frac{\chi}{v} \int d {\bf{r}} \hat{\rho}_p ({\bf{r}}) \hat{\rho}_s
({\bf{r}}) \right]
 \tag{1}
\end{align}
where $\int \hat{D} \{ {\bf{R}}_j \}$ denotes integration over all
chain configurations weighted by the Gaussian-chain statistics, and
$\int d {\bf{r}}_k$ denotes integration over the solvent degrees of
freedom. $\hat{\rho}_s (\bf{r})$ and $\hat{\rho}_p (\bf{r})$ are the
local instantaneous volume fraction of solvent and polymer,
respectively. The partition function Eq. 1 cannot be exactly
evaluated due to both the energetic interaction and the
incompressibility constraint.  We make the self-consistent field
(SCF) approximation, which involves (1) decoupling the interacting
system into a  noninteracting chains in fluctuating fields by a
identity transformation using functional integration over the
fluctuating fields, (2) replacing the functional integration over
the fluctuating fields by the saddle point approximation.  For
systematic derivation and numerical details we refer readers to the
standard literature\cite{Fredricksonbook,Fredrickson}. The free
energy of the system is then:
\begin{align}
\beta F_m = &\int d {\bf{r}} \frac{1}{v} \left[ \chi \rho_p
(1-\rho_p)-\omega_p \rho_p - \omega_s (1-\rho_p)  \right]\nonumber\\
& -m \ln Q_p  - e^{\beta \mu_s} Q_s +\ln\left(m!\right)\tag{2}
\end{align}
$Q_p$ is the single-chain partition function in the field$\omega_p$,
given by $Q_p=  \frac {1} {v} \int d {\bf{r}} q ({\bf{r}},N)$, where
$q({\bf{r}},\tau)$ is the chain propagator determined by the
diffusion equation:
$$\left[\frac{\partial}{\partial \tau} - \frac{b^2}{6} \nabla ^2 _{\bf{r}} +\omega_p \right] q({\bf{r}},\tau) =0 \eqno(3) $$
with initial condition $q({\bf{r}},0)=1$. $Q_s$ is the single
particle partition function of solvent in the field $\omega_s$,
given by $Q_s=\frac{1}{v} \int d{\bf{r}} \exp(-\omega_s)$. The
density profile $\phi_p (\bf{r})$ and the fields $\omega_p(\bf{r})$
and $\omega_s(\bf{r})$ are determined by the following
self-consistent equations:
$$\omega_p ({\bf{r}})-\omega_s ({\bf{r}})=\chi[1-2 \rho_p ({\bf{r}})] \eqno(4a)$$
$$\rho_p ({\bf{r}})= \frac{m }{Q_p} \int _0 ^{N} d \tau q({\bf{r}}, \tau) q({\bf{r}},N-\tau) \eqno(4b)$$
$$1-\rho_p ({\bf{r}})= e^{\beta \mu_s} \exp(-\omega_s ({\bf{r}})) \eqno(4c)$$
By solving equations 3 and 4 iteratively\cite{Discretize}, we obtain
the equilibrium density profile and free energy of a globule/cluster
in the subvolume.

\subsection{Effective Two-body Interaction}

If the polymer density in the subvolume is low, it is convenient to
integrate over the solvent degrees of freedom, which leads to
effective interaction between two polymer segments.  Applying the
identity transformation for $\hat{\rho}_s$ together with local
incompressibility, we recast the partition function in Eq. 1 into:
\begin{align}
\Xi  =& \frac{1}{m !  v ^{Nm}} \int D \omega_s \prod_{j=1}^{m} \int
\hat{D} \{ {\bf{R}}_j \}\nonumber\\
&\exp \left[ - \int d {\bf{r}} \frac{  1}{v} \left[ \chi
\hat{\rho}_p (1-\hat{\rho}_p)-\omega_s (1-\hat{\rho}_p)\right]+
e^{\beta \mu_s} Q_s \right] \tag{5}
\end{align}
Because of the high volume fraction of the solvent,  the fluctuation
of the field $\omega_s$ is small, and the functional integration
over $\omega_s$ in Eq 5 can be replaced by the saddle point value,
which yields
$$ \Xi  = \frac{1}{m !  v ^{Nm}}  \prod_{j=1}^{m} \int \hat{D} \{ {\bf{R}}_j \} \exp \left[-\beta H \right]$$
where $H$ is the Hamiltonian in the form of
\begin{align}
\beta H= \int d {\bf{r}} &\frac{1}{v} \left[ \chi \hat{\rho}_p
(1-\hat{\rho}_p)+ (1-\hat{\rho}_p) (\ln(1-\hat{\rho}_p)-1) \right]\nonumber\\
&-\frac{\beta \mu_s}{v} (1-\hat{\rho}_p)  \tag{6}
\end{align}
  Taking the
reference such that $H$ is zero in the reservoir of pure solvent
outside the subvolume, we obtain $\beta \mu_s=-1$.  Eq 6 can then be
written in the form of viral expansion in terms of instantaneous
volume fraction of polymer segments as
$$\beta H=\int d {\bf{r}} \frac{1}{v} \left[ (\chi -1) \hat{\rho}_p +\left( \frac{1}{2} - \chi \right) \hat{\rho}_p^2 +\frac{1}{6} \hat{\rho}_p^3+\cdots  \right] \eqno(7)$$
from which we identify the effective two-body interaction as
$\frac{1}{2}-\chi$. Although this is the identical expression of the
effective two-body interaction commonly used in the literature,  the
correspondence between the effective two-body interaction and the
F-H $\chi$ parameter in the previous
theories\cite{Florybook,deGennesbook,Grossbergbook} was constructed
for the {\it homogeneous, bulk} polymer solution; to our knowledge,
this connection has not been established at the level of
single-chain Hamiltonian. Here we have provided an explicit
demonstration, based on a saddle-point approximation which amounts
to neglecting the concentration fluctuation of the solvent, that the
expression is valid at the single-chain level.   If one further
includes the short-length-scale concentration fluctuations, the
$\chi$ parameter should more properly be interpreted as the
effective interaction parameter instead of the bare interaction
parameter\cite{WangJCP,Morse}. Renormalization group
theory\cite{Freedbook} showed that the second viral coefficient of
the polymer solution has corrections from the three-body
interaction. Grosberg and Kuznetsov\cite{Grosberg1} pointed out that
this correction vanishes in the limit of infinite chain length.
Therefore, both the effective two-body interaction and the second
viral coefficient vanish at $\chi_\Theta=\frac{1}{2}$ for infinitely
long chain.

\subsection{Phase Equilibrium}

The density profile and free energy of the globule and clusters
obtained from SCFT can be included into the framework of dilute
solution thermodynamics to reconstruct the bulk dilute polymer
solution. The free energy density of the entire solution with volume
$V_{t}$, including the translational entropy of clusters, can be
written as
$$\beta F/V_t=\sum _{m=1} ^{\infty} \left[ C_m \beta F_m + C_m (\ln \phi_m -1) \right] \eqno(8)$$
where $C_m$ and $\phi_m$ are respectively the concentration and
volume fraction of the cluster with association number $m$ (called
$m$-cluster henceforth) in the solution. $F_m$ is the free energy of
the $m$-cluster calculated by SCFT (obtained as an excess free
energy in the subvolume with respect to the pure solvent). In Eq 8,
we ignore the interaction between different clusters on the
assumption that the solution is sufficiently dilute. The equilibrium
concentration of $m$-clusters can be obtained by minimization of the
free energy density in Eq 8 subject to fixed total polymer
concentration $\sum_{m=1}^{\infty}mC_m$, which results in the
following cluster distribution:
$$\phi_m = (\phi_1)^m \exp (-\beta \Delta F_m) \eqno(9)$$
where $\Delta F_m \equiv F_m - mF_1$ is the free energy of formation
of the $m$-cluster from m single chain globule.

To study the coexistence between the polymer-poor phase and the
polymer-rich phase, we use Eq 8 with Eq. 9 to account for the large
localized density fluctuation in the polymer-poor solution, whereas
the polymer-rich phase is described by the F-H theory.  The phase
boundary is determined by the respective equality of the chemical
potential of the polymer and the solvent in the two coexisting
phases, which results in
\begin{align}
-\sum_{m=1}^{\infty} \frac{\phi_m}{mN} =&  \left(1- \frac{1}{N} \right) \phi_H + \ln(1- \phi_H) + \chi \phi_H ^2 \tag{10a}\\
\beta F_1 + \ln \phi_1 =& \ln \phi_H -1 + (1-N) (1-\phi_H) \nonumber\\
& + \chi N (1- \phi_H)^2 \tag{10b}
\end{align}
where $\phi_H$ is the equilibrium volume fraction of polymers in the
polymer-rich phase and $\phi_m$ is the equilibrium volume fraction
of the $m$-cluster in the dilute phase given by Eq 9.  The total
volume fraction of polymers in the dilute phase, $\phi_L$, is given
by $\phi_L=\sum_{m=1}^{\infty}\phi_m$.

\section{Globule Structure}
\label{sec:level2}

In the poor solvent, a single polymer chain adopts a compact
globular structure due to the unfavorable interaction between the
polymer segments and the solvent. The increase in local segment
density is countered by excluded volume (modeled by the
incompressibility in our theory) which prevents the globule from
collapsing into unbounded high density. Figure 1 shows the density
profile of a globule for different values of $\chi$ calculated by
SCFT.  Clearly the globule can be divided into a core region with
uniform density (denoted by $\rho_0$) and a surface region. The
globule core resembles a liquid droplet that contains a number of
uncorrelated parts of the chain.  By taking uniform fields
$\omega_p$ and $\omega_s$ in equations 2-4, we obtain the free
energy in the globule core to be
$$\beta F_c = \frac {N}{\rho_0} \left[ \chi \rho_0 (1-\rho_0) + (1-\rho_0)\ln (1-\rho_0) \right] \eqno(11)$$
which is the Flory-Huggins free energy without the translational
entropy of the polymer due to the fixed center of mass. In the limit
of large $N$, the Laplace pressure due to curvature can be
neglected, and the polymer density of the globule core can be
obtained approximately by balancing the osmotic pressure inside the
globule with that of the pure solvent outside ($\approx 0$), which
yields $\rho_0$ as the nontrivial root of the following
equation\cite{Eq2}
$$\rho_0+ \ln(1-\rho_0)+ \chi \rho_0^2  =0 \eqno(12)$$
For $\chi$ close to 0.5, Eq 12 has the symptomatic solution of
$\rho_0 \approx 3 (\chi -0.5)$. As shown in Figure 1, the core
density becomes lower with the decreasing $\chi$, which increases
the size of the globule as a result of swelling.  At the same time,
the interfacial region becomes thicker and more diffuse. We see from
Figure 2 that the thickness of the interfacial region scales with
the Kuhn length $b$ as pointed out by Grosberg and
Khokhlov\cite{Grossbergbook} and is independent of the chain
stiffness parameter $p=b^3/v$.  Because the dependence on $p$ is
relatively straightforward, in most of the subsequent discussions,
we choose a specific $p$ corresponding to $v=4 \pi b^3/3$.

\begin{figure}[h]
\centering
\includegraphics[width=0.5\textwidth]{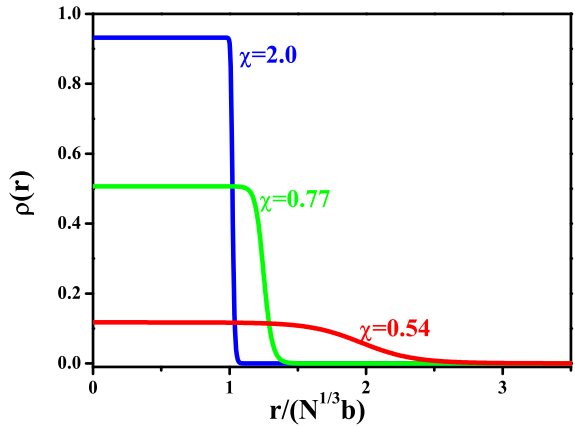}
\caption{Density profile of a globule for three values of $\chi$
with $N=10^4$  and $v=4\pi b^3/3$. $\rho(r)$ is the local volume
fraction of polymer with $r$ the radial axis starting from the
globule center. \label {1}}
\end{figure}

\begin{figure}[htbp]
\centering
\includegraphics[width=0.5\textwidth]{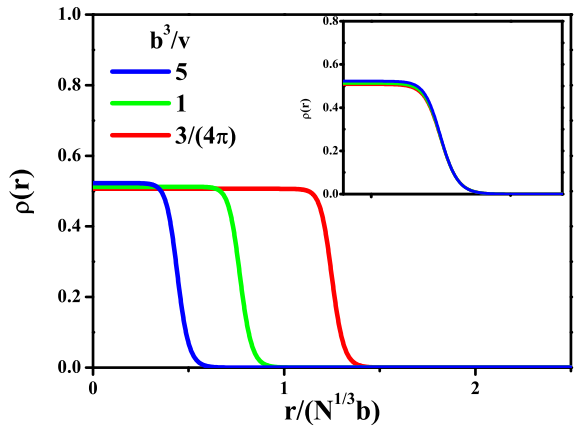}
\caption{Effect of the chain stiffness parameter $p=b^3/v$ on the
density profile of a globule with the same Kuhn length. $\chi=0.77$
and $N=10^4$. The insert shows the overlap of the interfacial region
when the density profiles are shifted. \label {2}}

\end{figure}

Based on the density profile, we can define the radius of gyration
($R_g$), the core radius ($R_0$) and the surface diffuseness ($h$)
to characterize the globule size, which are given as:
\begin{align}
&R_g =\left( \frac{\int_0 ^\infty r^2 \rho(r) 4\pi r^2 dr}{\int_0 ^\infty \rho(r) 4\pi r^2 dr} \right)^{1/2}\tag{13a}\\
&R_0 =(3/5)^{1/2} \left[ 3Nv/ (4 \pi \rho_0) \right]^{1/3}\tag{13b}\\
&h =R_g-R_0 \tag{13c}
\end{align}
$R_0$ is the radius of gyration for a uniform sphere with polymer
density $\rho_0$ that contains the same total amount of chain
segments as in the globule; in other words, $r=(5/3)^{1/2}R_0$ is
the location of the Gibbs dividing surface. $h$ describes the
deviation of the globule from the uniform sphere, thus
characterizing the diffuseness of the globule surface. It should be
noted that $h$ is not a linear measure of the true thickness of the
globule surface (denoted by $h_T$) within which the polymer density
changes from $\rho_0$ to 0.  Dobrynin and Rubinstein\cite{Dobrynin}
pointed out that the surface thickness of the globule has a length
scale of one thermal blob $h_T \sim b/ \rho_0$.  For a globule with
a large uniform core (i.e., $h_T/R_0 \ll 1$), it can be shown that
the surface diffuseness scales as $h \sim h_T^2 /
R_0$.\cite{Surfacediffuseness} In this work, we choose $h$ rather
than $h_T$ to characterize the globule surface because $h$ is
defined based on $R_g$, which is an experimentally measurable
quantity.

With the definition of $R_0$ and $h$, we can separately examine the
volume contribution and the surface contribution to the globule
size. The core density and the globule size as a function of $\chi$
for polymer with $N=10^4$ are shown in Figure 3.  When $\chi$ is
well above $0.5$, $R_g$ coincides with $R_0$ while $h$ remains very
small and is negligible compared to $R_0$. The globule surface is of
the thickness of one thermal blob ($h_T \sim b/\rho_0$, $h \sim b
\rho_0^{-5/3}N^{-1/3}$), and the entire globule is space-filled by
$\sim \rho_0 ^2 N$ thermal blobs. In this regime, the globule can be
envisioned as a uniform sphere: the increase of the globule size as
$\chi$ decreases is due to the swelling of the globule core (see
Figure 3b), given by $R_g \approx R_0 \sim (Nv/\rho_0)^{1/3}$. As
$\chi$ approaches $0.5$, $R_g$ increases rapidly and deviates from
$R_0$. At the same time, $h$ increases sharply and becomes
comparable to $R_0$. The thermal blob expands, reaching a size
comparable to the whole globule.  The rapid increase of the globule
size in this regime indicates melting of the globule, which takes
place by both the swelling of the globule core and the widening of
the interface.

The $\chi$ dependence of the globule size and core density shown in
Figure 3 suggests that the single chain conformation prior to the
globule-to-coil transition involves three states: the
\emph{fully-collapsed globule}, the \emph{swollen globule} and the
\emph{molten globule}, in the order of decreasing $\chi$. For large
$\chi$ ($\chi
> 1$), the globule is
fully-collapsed, with core density close to one and a sharp
interface: $\rho_0 \approx 1$, $h_T / R_0 \sim N^{-1/3}$ and $h /
R_0 \sim N^{-2/3}$. The fully-collapsed globule can further
transform to the ordered solid-globule as pointed out by Zhou et
al.\cite{Zhou}, which is out of the scope of this work.  In the
intermediate regime, where $\chi-0.5\ll1$ but $(\chi - \chi_\Theta)
N^{1/2} \gg 1 $, the globule is swollen. The core density is
significantly lower than one, while the interfacial thickness is
still narrow compared to the core radius: $\rho_0 << 1$, $h_T / R_0
\sim \left[(\chi - \chi_\Theta) N^{1/2}\right]^{-2/3}$, $h / R_0
\sim \left[(\chi - \chi_\Theta) N^{1/2}\right]^{-4/3}$. Finally in
the $\Theta$ regime, where $(\chi - \chi_\Theta) N^{1/2} \sim O(1)$
or less, the globule is molten, with a very dilute core, comparable
to a Gaussian coil, and an interfacial thickness comparable to the
core radius: $\rho_0 \sim (\chi- \chi_\Theta) \sim N^{-1/2}$, $h_T/
R_0 \sim O(1)$, and $h/ R_0 \sim O(1)$.
\begin{figure}[htbp]
\centering
\includegraphics[width=0.45\textwidth]{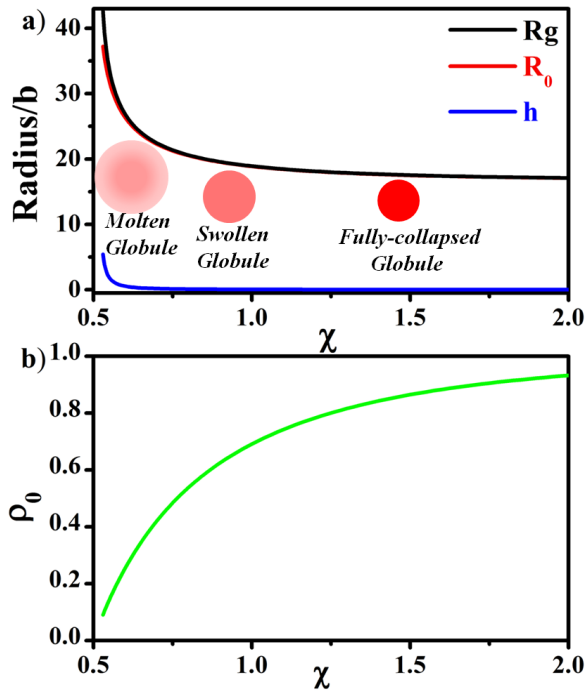}
\caption{a) Radius of gyration ($R_g$), core radius ($R_0$) and
interfacial diffuseness ($h$) and b) core density($\rho_0$) of the
globule as a function of $\chi$ with $N=10^4$. The schematic
inserted in a) represents the three states in globule structure: the
\emph{fully-collapse globule}, the \emph{swollen globule} and the
\emph{molten globule}. \label {1}}
\end{figure}

That the globule structure may involve more than one states has been
suggested before. Wu and Wang\cite{Wu1,Wu2} proposed that the
globule structure involves two states: the fully-collapsed globule
and the molten globule, based on the consideration that the ratio
between the radius of gyration and hydrodynamic radius for the
fully-collapsed globule is $(3/5)^{1/2}$ whereas this ratio for the
molten globule is smaller than $(3/5)^{1/2}$ (this criterion is
equivalent to whether the globule has a sharp or diffuse surface).
However, the ``fully-collapsed'' globule observed in their
experiment contains $66\%$ solvent inside the globule volume, which
is more properly considered as the swollen globule in our
definition.  If we choose $\chi=1$ as the criterion for separating
the swollen globule from the fully-collapsed globule, then the range
of the swollen globule is roughly between $\chi_\Theta$ and $\chi=1$
for large $N$; this can correspond to a large temperature window in
experiments. The swollen state for the infinitely long chain
persists all the way to the $\Theta$ point, as we discuss below.

It is instructive to note an analogy between the swollen globule and
the semidilute polymer solution in good solvent.  The existence of
both these states is due to the long chain length.  Indeed the
molten globule, the swollen globule and the fully-collapsed globule
can be considered the respective analogues of the dilute, semi
dilute, and concentrated solutions.  Just as there is no clear-cut
boundary between the semidilute and concentrated solutions, the
distinction between a swollen globule and fully collapse globule is
not a sharp one.  And just as the concentration range for the dilute
regime becomes narrower as the chain length increases, so the region
of the molten globule shrinks with increasing chain length.
Likewise, with decreasing chain length, the window for the existence
of the swollen globule shrinks, and eventually disappears for very
short chains.

We should emphasize that the density profile of the globule is
calculated using the flexible Gaussian chain model. If the thickness
of the globule surface becomes comparable to the Kuhn length, which
happens for very large $\chi$, the chain segments can become
oriented by the interface.  In that case, the flexible Gaussian
chain is no longer a valid description and chain rigidity and local
nematic order may have to be accounted for
explicitly\cite{Szleifer}.  However, we find that for $\chi=1.0$
(with the core density $\rho_0=0.7$) the interfacial width is
$h_T=5b$ and is $h_T=10b$ for $\chi=0.77$ (with the core density
$\rho_0=0.52$). The interfacial width decreases to the Kuhn length
at about $\chi=2.0$, corresponding to a core density $\rho_0=0.93$.
Therefore, there is a significant range of parameter space in which
the Gaussian chain model remains valid.

\section{Determining $\chi_\Theta$ from the single-chain globule}
\label{sec:level3}

The globule structure obtained by our SCFT calculation is valid
under the condition that the surface energy of the globule is much
larger than $kT$; otherwise shape fluctuation must be taken into
account and the assumption of spherical symmetry becomes
questionable. This condition is equivalent to the requirement that
the number of thermal blobs $\rho_0 ^2 N$ must be large, i.e., $
\rho_0 ^2 N \gg 1$.  To satisfy this requirement, the globule state
should be within the parameter space $(\chi - \chi_\Theta)N^{1/2}
\gg 1$, i.e., the chain should be in the fully-collapsed or swollen
globule.  (Equation 17b provides a more quantitative Ginzburg
criterion.) This requirement would at first seem to prevent us from
applying the theory to the close vicinity of the $\Theta$ point.
However,  as the chain length increases, the width of the $\Theta$
region (in $\chi$) shrinks and the transition becomes increasingly
sharper\cite{deGennes,Grosberg1,WQ}. In other words, the state of
the swollen globule persists closer to the $\Theta$ point. In the
limit of infinitely long chain, the swollen globule persists all the
way to the $\Theta$ point. Therefore, the chain length dependence of
the melting of the swollen globule can be used to probe the width of
the $\Theta$ region and the location of the $\Theta$ point.

\begin{figure}[htbp]
\centering
\includegraphics[width=0.5\textwidth]{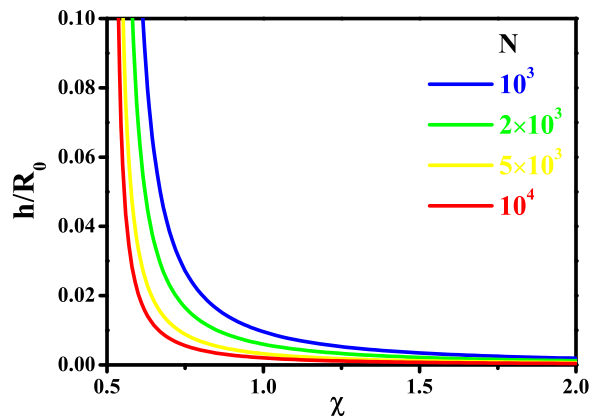}
\caption{The ratio between the surface diffuseness and the core
radius ($h/R_0$) as a function of $\chi$ for different chain length.
\label {1}}
\end{figure}

Figure 4 shows the ratio between the surface thickness and the core
radius of the globule ($h/R_0$) as a function of $\chi$ for several
different chain lengths.  As $\chi$ decreases close to 0.5, $h/R_0$
exhibits a rapid rise as a result of the melting of the globule;
this rise is sharper for larger $N$.  From Figure 4, we can define
the onset of the melting (denoted by $\chi_N(\lambda)$) as the value
of $\chi$ at which $h/R_0$ reaches a threshold $\lambda$.
$\chi_N(\lambda)$ thus represents the onset of the globule-to-coil
transition as well as the boundary of the $\Theta$ region on the
globule side.  Clearly, $\chi_N(\lambda)$ depends on chain length.
Figure 5 shows the linear relation of $\chi_N(\lambda)$ versus
$N^{-1/2}$ for three different levels of $\lambda$; this linearity
demonstrates that the width of the $\Theta$ region is proportional
to $N^{-1/2}$, in agreement with the results from the scaling
argument\cite{deGennes} and simulation\cite{Binder}.

\begin{figure}[htbp]
\centering
\includegraphics[width=0.5\textwidth]{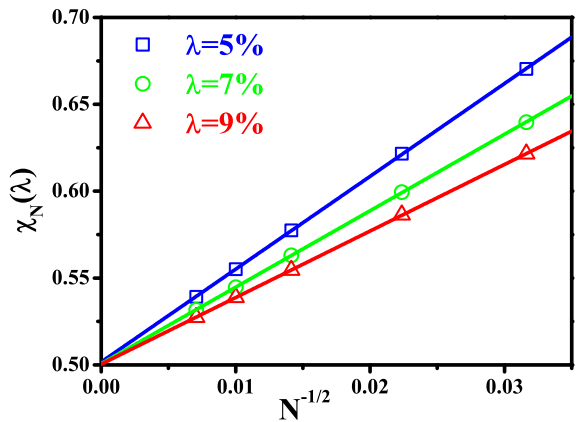}
\caption{The onset of globule melting $\chi_N(\lambda)$ plotted vs
$N^{-1/2}$ for three different $\lambda$. Points represent the
results calculated by SCFT. Straight lines are the linear fit to the
points. Three lines intersect to a common point as $N \to \infty$,
which yields $\chi_\Theta=0.5$. \label {1}}
\end{figure}

As $N$ increases, $\chi_N(\lambda)$ for the different choices of the
threshold $\lambda$ become closer, indicating that the transition
becomes sharper for longer polymer chains. Extrapolating
$\chi_N(\lambda)$ versus $N^{-1/2}$ to the limit of $N \rightarrow
\infty$, $\chi_N(\lambda)$ for different $\lambda$'s converge to a
common point that is independent of the choice of $\lambda$,
indicating that the transition is infinitely sharp as $N \to
\infty$. From the perspective of single-chain conformation, $\Theta$
point is defined as the abrupt change in chain size from globule to
coil for an infinitely long chain. The common intersection shown in
Figure 5 yields $\chi_\Theta=0.50$.  This value coincides with the
result of F-H theory and is also consistent with the vanishing of
the second viral coefficient derived in Sec II.  We comment that the
coincidence of $\chi_\Theta=0.50$ with the prediction of the F-H
theory is not a trivial result, nor does it mean that the F-H theory
itself is valid.  Indeed, the F-H theory is a poor description of
the dilute solution, including the solution near the $\Theta$ point.
However, if we focus on the core of an isolated globule, it is a
homogeneous region containing many uncorrelated blobs of the chain,
sharing the same physical picture as the F-H theory.  In the limit
of the infinite chain length, the swollen globule state persists all
the way to the $\Theta$ point where the effective two-body
interaction vanishes.

\begin{figure}[htbp]
\centering
\includegraphics[width=0.5\textwidth]{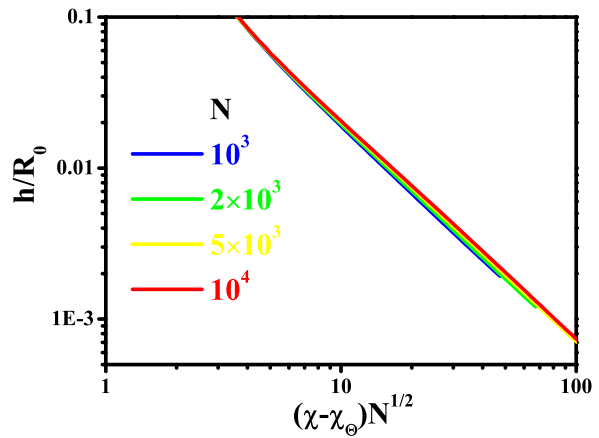}
\caption{Universal behavior in the scaling region. log-log plot of
$h/R_0$ vs $\left(\chi - \chi_\Theta \right)N^{1/2}$. \label {1}}
\end{figure}

Using $\chi_\Theta$, $h/R_0$ can now be plotted as a function of the
scaling variable $\left(\chi - \chi_\Theta \right)N^{1/2}$. Figure 6
shows that $h/R_0$ for different chain lengths collapse onto each
when plotted against $\left(\chi - \chi_\Theta \right)N^{1/2} $,
confirming the universal behavior near the $\Theta$ point:
$$ h/R_0=f\left[ \left(\chi - \chi_\Theta \right)N^{1/2} \right]   \eqno(14)$$
where $f$ is a universal function. In the range $\left(\chi -
\chi_\Theta \right)N^{1/2}>10$ (where the globule is in the swollen
state with low core density), the slope of the log-log plot is
$-1.4$, close to the exponent $-4/3$ we have obtained in Sec III by
using the scaling argument. The universal behavior of the single
chain conformation around the $\Theta$ point, initially predicted by
de Gennes\cite{deGennes}, is usually difficult to achieve in the
globule state in computer simulation\cite{Binder} due to the
limitation of chain length.  For short chains, the swollen globule
either is not well-defined or cannot persist to a very dilute core,
thus making the window for observing the scaling regime unclear or
too small. Here, by applying SCFT to long chains ($N>10^3$), we are
able to more definitively examine this scaling regime in the globule
state.

\section{Determining $\chi_\Theta$ from solution thermodynamics}
\label{sec:level3}

The single chain structure will of course affect the solution
properties. The compaction of chain segments into a globule
significantly reduces the polymer-solvent contact, which leads to
enhancement of the solubility of polymer in poor solvent by several
orders of magnitude compared to the prediction of the F-H theory as
shown in our earlier paper\cite{Wang1}. The governing equation for
the phase equilibrium of polymers in poor solvent is given by Eq 10
in Sec IIC. In Eq 10a, the osmotic pressure for long polymer chain
in the dilute solution is very small ($\sum_{m=1}^{\infty} \phi_m
/mN\approx 0$). Thus, from the similarity between Eq 10a and Eq 12,
we obtain for the volume fraction of the polymer-rich phase,
$$\phi_H \approx \rho_0 \eqno(15)$$
On the other hand, based on our analysis of the globule structure in
Sec III, the free energy of the single chain globule in Eq 10b can
be divided into the volume contribution and the surface contribution
as $F_1=F_c + F_{s}$. It can be shown that this definition is
consistent with the definition based on the monomeric osmotic
pressure across the interface used in Ref. \cite{Grossbergbook}. By
substituting $F_c$ in the form of Eq 11 into Eq 10b, and making use
of Eq 12 and Eq 15,  we obtain the relation between the equilibrium
volume fraction of polymers in the dilute phase and the surface
energy of the globule
$$\ln \phi_L \approx -\beta F_{s} +\ln \rho_0 - \rho_0 \ = - \gamma A +\ln \rho_0 - \rho_0 \eqno(16)$$
from which we identify the surface tension $\gamma$.
$A=(36\pi)^{1/3} (Nv/\rho_0)^{2/3}$ is the surface area of the
globule.  Noting that the natural dimension of the surface tension
is $b/v$ ($kT$ is taken to be 1), a dimensionless surface tension
can be defined as $\gamma v^{2/3}$ which thus contains an overall
$p^{1/3}$ dependence; this is the origin of the $p^{1/3}$ in Eqs.
17a and 17b. Eq 16 indicates that, the excess free energy of a
globule is due to its surface energy, and not due to (uniform)
mixing of polymer segments and solvents assumed in the F-H theory.
Equations 15 and 16 reveal the intrinsic connection between the
macroscopic phase behavior of the polymer solution and the
microscopic structure and property of the single globule.  This
connection also allow us to extract information of single chain
globule from solution thermodynamics.

\begin{figure}[htbp]
\centering
\includegraphics[width=0.5\textwidth]{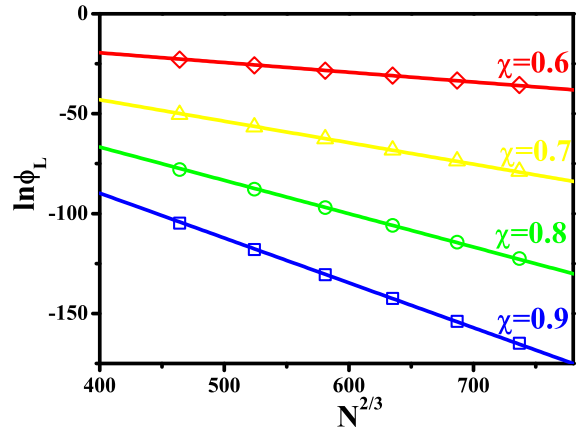}
\caption{Linear relation between $\ln \phi_L$ and $N^{2/3}$. Points
represent the numerical results based on Eq 10. Straight lines are
the linear fit of the points. \label {1}}
\end{figure}

The linear relation between $\ln \phi_L$ and $N^{2/3}$ is confirmed
by numerically solving Eq. 10 as shown in Figure 7. From the slope
of $\ln \phi_L$ versus $N^{2/3}$,  we can extract the surface
tension of a single globule. Figure 8 shows that for $\chi$ much
larger than $\chi_\Theta$, $\gamma$ decreases linearly with $\chi$
as $\gamma v^{2/3}=0.11\chi -0.07$. This linear dependence on $\chi$
agrees with the result of Weber and Helfand for the interfacial
tension of a planar interface\cite{Helfand}.  As $\chi$ decreases,
$\gamma$ deviates from the linear behavior and approaches zero. The
disappearance of the surface tension of the globule serves as
another definition of the $\Theta$ point.  By extrapolating the
numerical data to the limit of $\gamma=0$, Figure 8 yields
$\chi_\Theta = 0.50$, which coincides with previous results obtained
from other definitions. Figure 8 also shows that, $\gamma$ follows a
quadratic form $\gamma v^{2/3}= 0.52 (\chi-0.5)^2$ for $\chi -
\chi_{\Theta} \ll 1$. The quadratic form of $\gamma$ near the
$\Theta$ point is in agreement with scaling prediction by Lifshitz
et al.\cite{Lifshitz}.

\begin{figure}[htbp]
\centering
\includegraphics[width=0.5\textwidth]{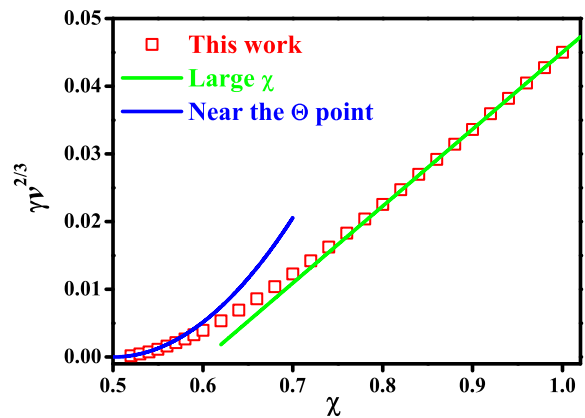}
\caption{Dimensionless surface tension of the globule $\gamma
v^{2/3}$ as a function of $\chi$. Extrapolating the numerical data
to $\gamma=0$ reveals $\chi_\Theta=0.5$. $\gamma$ is fitted linearly
for large $\chi$ as $\gamma v^{2/3} = 0.11 \chi -0.07$, and fitted
by a quadratic function $0.52(\chi-0.5)^2$ near the $\Theta$ point.
\label {1}}
\end{figure}

In the framework of F-H theory, des Cloizeaux and Jannink obtained
$\phi_L=\frac{3}{e} \left(\chi - \frac{1}{2} \right) \exp \left[-
\frac{3}{2} \left( \chi - \frac{1}{2}  \right)^2 N \right]$. By
substituting the expression of $\gamma$ and $\rho_0$ into Eq 16, our
theory yields the equilibrium volume fraction of polymers in the
dilute phase as
\begin{align}
& \phi_L =\frac{1}{e} \exp\left[ -p^{1/3}(0.89\chi -0.53)N^{2/3} \right]\nonumber\\
& for \; large \; \chi \tag{17a}\\
& \phi_L=3(\chi - 0.5) \exp\left[ -1.93p^{1/3}(\chi-0.5)^{4/3}
N^{2/3} \right]\nonumber\\
 & for \; \chi-\chi_\Theta \ll 1
\tag{17b}
\end{align}
where we have inserted the dependence on the stiffness parameter
$p=b^3/v$.  We note that an equation similar to Eq 17b has been
given in Ref. \cite{Grossbergbook} (Eq. 26.8). However, our Eq. 17b
provides the numerical prefactor (1.93) in the scaling dependence in
the exponential, which was not provided in Ref.
\cite{Grossbergbook}. Eq. 17a is a completely new result. These
equations show very different scaling behavior of the solubility of
polymers in poor solvent predicted by our theory compared to the F-H
theory\cite{Cloizeaux}, both in the $\chi$ dependence and in the $N$
dependence.

The exponent in Eq. 17b can be considered as a quantitative Ginzburg
criterion for the validity of applying the SCFT, i.e., SCFT is
applicable if $p^{1/3}(\chi-0.5)^{4/3} N^{2/3} > 1$.  Thus
increasing the stiffness parameter $p$ allows a closer approach to
the $\Theta$ point.  This conclusion is consistent with results from
computer simulation by Withers et al. \cite{Withers}.

From Eq 17b, we obtain the slope of the phase boundary near the
$\Theta$ point as

\begin{align}
\frac{\partial \phi_L}{\partial \chi} =& \left[ 3 -7.74 p^{1/3}(\chi
-
0.5)^{4/3}N^{2/3} \right] \nonumber\\
& \exp \left[ -1.93p^{1/3} (\chi - 0.5)^{4/3}N^{2/3} \right]
\tag{18}
\end{align}
Consistent with the validity of the SCFT, Eq 18 is valid
in the swollen globule state, which can persist all the way to the
$\Theta$ point for infinitely long chain. In the limit of $N \to
\infty$, $\partial \phi_L /
\partial \chi=0$ if $\chi > 0.5$. However, $\partial
\phi_L /
\partial \chi$ jumps to a finite value at $\chi = 0.5$, which suggests that $\chi=0.5$ is the critical point in the phase diagram of a polymer
solution in the limit of infinite chain length\cite{Footnote}. This
establishes the $\Theta$ point defined from the perspective of
solution thermodynamics.

These results confirm that the $\Theta$ point defined as the abrupt
change in chain size from globule to coil and as the critical point
in the phase diagram are consistent with each other in the limit of
infinite chain length.  This consistency reveals the intrinsic
connection between the single chain structure and its solution
properties.  From the perspective of solution thermodynamics our
results suggest an alternative experimental determination of the
$\Theta$ point from measuring the slope of $\ln \phi_L$ versus
$N^{2/3}$ at different temperatures.  Since this approach is based
on finite polymer concentrations, it may be more easily conducted
than the single-chain measurement.

\section{Conclusions}
\label{sec:level3}

In this work, we have presented a theory that unifies the study of
the single-chain structure and solution thermodynamics for polymers
in poor solvents, using the language of the Flory-Huggins parameter.
Our theory captures the large localized density fluctuation in
dilute polymer solutions and the change in chain conformation with
the solvent quality that are missing in the F-H theory. The
structure of a globule is studied by SCFT, which affords a more
accurate description of the density profile of the globule and its
free energy for finite chain lengths. By relaxing the assumption of
the virial expansion of the local interaction as assumed in previous
theories, our theory can be applied to globules with relatively high
monomer density, which facilitates the calculation of the surface
tension of the globule and the solubility of polymers in dilute
solution for large values of $\chi$.  On the other hand, by avoiding
the ground state dominance approximation, the SCFT is capable to
describe globules with diffuse interfaces.  These advantages allow
us to study the globule structure and the thermodynamics of the
globule melting in a unified theoretical framework. The chain-length
dependence of the globule melting provides the essential information
on the approach to the $\Theta$ point.

We briefly summarize the key new results of this work. First, we
show that the chain conformation involves three globular states
prior to the globule-to-coil transition: the \emph{fully-collapsed
globule}, the \emph{swollen globule} and the \emph{molten globule};
this identification clarifies the ambiguity in the experimental
studies of the globule structure.  Second, we provide numerical
verification of the universal behavior as a function of the scaling
variable $(\chi - \chi_\Theta)N^{1/2}$ near the $\Theta$ point,
which has not been achieved by previous computer simulation in the
globule state due to the limitation of the chain length. Third, we
provide new results for the solubility of the polymers in dilute
solution compared to the results of des Cloizeaux and Jannink based
on the F-H theory.  The large $\chi$ result is completely new while
the result near the $\Theta$ point provides the missing numerical
prefactor in previous work.  Fourth, we demonstrate the consistency
of the $\Theta$ points defined by the different criteria in the
limit of infinite chain length: the disappearance of the second
viral coefficient in the dilute polymer solution, the abrupt change
in chain size from globule to coil and the critical point in the
phase diagram of polymer solution. Fifth, we find $\chi_\Theta =0.5$
from all three different criteria, which \emph{coincides} with the
prediction of F-H theory.  Lastly, although the expression itself is
known and widely used in the literature, we provide the explicit
derivation that shows the effective two-body interaction in the {\it
single-chain Hamiltonian} is given by $0.5-\chi$.

Although $\chi_\Theta=0.5$ can be directly obtained through the
viral expansion of the local Hamiltonian if the polymer-solvent
interaction is parameterized by the Flory $\chi$ parameter, the two
methods we present in this paper, that is tracking the chain-length
dependence of the melting of the globule and tracking the
chain-length dependence of the solubility of polymer solution, are
more general approaches in determining the $\Theta$ point.  For
example, the approaches can also be used for heteropolymers.
Moreover, the consistency of the $\Theta$ point defined by the
different criteria allows us to choose a convenient way to measure
the $\Theta$ point in  experiment.  Our results suggest that the
$\Theta$ point can be determined by measuring the slope of $\ln
\phi_L$ versus $N^{2/3}$ for different temperatures, which is a
finite-concentration instead of a single-chain measurement.


\begin{thebibliography}{}
\bibitem{Eq2}
We note that in our previous work \cite{Wang1}, the equation (Eq. 2
in that reference) incorrectly contains an extra $-N^{-1} \rho_0$
term.  However, that equation was not used for any of the actual
calculations, so the results and conclusions in Ref. \cite{Wang1}
are unaffected by this error.
\bibitem{Surfacediffuseness}
This can be easily shown by assuming the density profile to be a
flat region plus a decaying function of range $h_T$.
\bibitem{Footnote}
Strictly speaking, the critical point corresponds to $\partial
\phi_L /
\partial \chi \rightarrow \infty$.  However, since our method is only valid in the swollen globule regime, we are unable to reach the critical point.
Therefore, rather than a divergence, we obtain a discontinuity.
\end{thebibliography}
\end{document}